\documentstyle[prd,aps,floats]{revtex}

\newcommand{\be}{\begin{equation}}
\newcommand{\ee}{\end{equation}}
\newcommand{\bea}{\begin{eqnarray}}
\newcommand{\beas}{\begin{eqnarray*}}
\newcommand{\eea}{\end{eqnarray}}
\newcommand{\eeas}{\end{eqnarray*}}
\newcommand{\ba}{\begin{array}}
\newcommand{\ea}{\end{array}}
\newcommand{\bi}{\begin{itemize}}
\newcommand{\ei}{\end{itemize}}
\newcommand{\ben}{\begin{enumerate}}
\newcommand{\een}{\end{enumerate}}

\begin{document}
\draft

\input epsf \renewcommand{\topfraction}{0.8} 
\twocolumn[\hsize\textwidth\columnwidth\hsize\csname 
@twocolumnfalse\endcsname

\title{Assisted inflation via tachyon condensation}
\author{Anupam  Mazumdar$^a$, Sudhakar Panda$^{b,a}$ and 
Abdel P\'erez-Lorenzana$^{a,c}$}
\address{
$^a$ The Abdus Salam International Centre for Theoretical Physics, I-34100,
Trieste, Italy \\
$^b$ Harish Chandra Research Institute, Chhatnag Road, Jhoosi, Allahabad-211019, 
India\\
$^c$ Departamento de F\'{\i}sica,
Centro de Investigaci\'on y de Estudios Avanzados del I.P.N.\\
Apdo. Post. 14-740, 07000, M\'exico, D.F., M\'exico}
\maketitle

\begin{abstract}
In this paper we propose a new mechanism of inflating the Universe with
non-BPS $D4$ branes which decay into stable $D3$ branes via
tachyon condensation. In a single brane scenario the tachyon potential is very
steep and unable to support inflation. However if the universe lives in a stack
of branes produced by a set of non-interacting unstable $\tilde {D4}$ branes,
then the associated set of tachyons may drive inflation along our $3$ spatial
dimensions.  After tachyon condensation the Universe is imagined 
to be filled with a set of parallel stable $D3$ branes.
We study the scalar density perturbations and reheating within this setup.
\end{abstract}

\vspace*{-0.6truecm}
\vskip2pc]


\section{Introduction} 

Inflation \cite{inf} is perhaps the only known mechanism 
which can dynamically solve the 
flatness and the horizon problem of the Universe. The inflating field can also 
produce the density perturbations causally to match the startling observation 
made by the COBE satellite. The observation has measured the amplitude of the 
density perturbations in one part in $10^{5}$ with a slightly tilted 
spectrum \cite{bunn}. It is this result which has boosted the hope that the 
coming generation satellite experiments might be able to constrain or 
possibly rule out some of the models. It is a theoretical and 
experimental challenge to find an exact shape of the potential 
and lacking this leads to more than a dozen viable inflationary models 
with some common features.

The moduli of string theory might provide us a natural candidate as an inflaton
\cite{binetruy}. However, as a zero mode, moduli lacks potential with a stable minimum. 
It is not clear at the moment how non-perturbative effects would generate a potential 
for them and what could be the appropriate scale for inflation \cite{banks}.
The moduli also accompanies a host of problems for cosmology because of their 
Planck mass suppressed couplings to the matter fields \cite{brustein}. On the
other hand string theory also provides various solitonic states in the
non-perturbative spectrum. These solitonic states are strong-weak
(S-duality) coupling partners of the states in the perturbative spectrum of 
the dual string theory. These states are known as Dp branes \cite{polchi}, where
the open strings end on a hypersurface with $p$-spatial and one time-like
dimension. These objects couple to gauge fields (coming from the RR sector) of rank
$(p+1)$, where $p = -1,0,1,2,..,9$, and,  $p$ is odd (even) for Type IIB
(A) theory. These branes are known as BPS branes and are stable under variation of
the string coupling. On the contrary, string theory also admits non-BPS Dp-branes
(also labeled as $\tilde{Dp}$) 
for odd (even) $p$ in Type IIA (B) \cite{sen1}. Generically these are
unstable because of the presence of tachyonic state(s) in their world volume, 
though in some cases they may become stable. For the stable non-BPS branes the 
tachyonic states are projected out of the spectrum by orbifolding, or, 
orientifolding procedure \cite{sen2}. They can be the lightest states
in the spectrum carrying some conserved charges and hence cannot decay to anything
else and thus are stable objects. For unstable non-BPS branes where there are
tachyonic states living on the world volume of the brane the tachyon can condense
over a kink yielding a brane with one dimension less \cite{sen3}. For an
example, let us consider a coincident pair of Dp-brane and anti-Dp-brane 
(with opposite RR charge and both independently being BPS). There is a 
complex tachyon field $T$ living on the world volume of this system. If we 
integrate out all the massive modes on the world volume except the tachyon, we 
obtain the tachyon potential $V (T)$ with $T = 0$ being the maximum of the 
potential. Since, there is $U (1)\times U(1)$ gauge field living on
the world volume of the brane-antibrane system, the tachyon picks up a phase
under the gauge transformation of each of this $U (1)$, and finally the 
potential becomes a function of $|T|$. Thus, the minimum of the potential occurs at 
$T = T_0 e^{i\theta}$ for some fixed $T_0$ but arbitrary $\theta$. 
According to the conjecture of Sen, at the minimum the sum of the tension 
of the brane and the anti-brane, and, the (negative) potential energy of the 
tachyon should be exactly zero. Thus, the tachyonic ground state at the minimum of 
the potential is indistinguishable from the vacuum. However, instead of 
considering the tachyonic ground state if we consider a tachyonic kink 
solution, one finds that the energy density is concentrated around 
a $(p-1)$ dimensional subspace and the solution describes a
$(p-1)$-dimensional brane. 
However, this solution is not stable since the manifold describing the minimum
of the tachyon potential is a circle. This is also
consistent with the identification of the kink solution with a non-BPS brane of
a string theory since it has a tachyonic mode living on its own world volume.

One can continue one step further by considering a non-BPS $p$-brane. The tachyon
which lives on its world volume is a real tachyon. But there is a $Z_2$ symmetry on
the world volume of this non-BPS brane under which the tachyon changes sign. 
Therefore, in this case, $\pm T_0$ becomes the minimum of the tachyon potential 
obtained by integrating out other massive modes and at the minimum of the 
potential the sum of the (negative) potential and the tension of the 
non-BPS brane vanishes. Like in the previous case, instead of the tachyonic 
ground state if we consider the tachyonic kink solution, the configuration 
actually describes a $(p-1)$-brane. However, the kink solution now is a stable 
solution, since the manifold describing the minimum of the tachyon potential 
consists of just two disconnected points $\pm T_0$  (related by
a $Z_2$-symmetry). This configuration, thus is
identified with a BPS $(p-1)$-brane of the theory.

The tachyonic potential, either on the world volume of the brane-antibrane
system, or, on the world volume of the non-BPS brane is not calculable
explicitly. However, an approximate nature of the potential with the properties as
mentioned above has been conjectured and such a potential is well supported by string
field theory calculations. We will assume such expressions for the tachyon
potential on  either a single non-BPS $D4$-brane or a system of non
coincident $\tilde{D4}$-branes of Type IIB string theory and test whether
tachyons could  play any role in the early Universe.

Recently, there has been a spate of interesting ideas which propose the
Universe as a D3 brane, or a stack of  D3 branes on top of each other, where
ultimately the Standard Model fields are trapped \cite{extra}. These new ideas
come up with a host of new cosmological issues and several new ideas have been
discussed, such as  inflationary~\cite{lyth,lukas,abdel,alexander,burgess,gia},
and  non-inflationary scenarios \cite{turok,kallosh}. One of the  common
feature these models possess is to inflate the brane by altering  inter-brane
separation. From the four dimensional point of view this corresponds to a
scalar field known as radion. Fluctuations  in the radion field can also lead
to the density perturbations observed by  the COBE satellite and eventually the
radion couplings to matter fields lead to its decay to produce large entropy. A
completely new possibility may  arise where inflation does not depend on the
relative motion of the branes.  Inflation now depends on the intrinsic
properties of an unstable  $\tilde D4$ brane which decays into a stable $D3$
brane. This may happen due to the dynamics of a zero mode of the tachyon field.
Usually, the tachyon potential  is quite steep, and,  can not support
slow-rolling of the tachyon field. Instead, the tachyon rolls down extremely
fast. However, under certain circumstances it  is indeed possible to construct
a model where more than one tachyon fields are present  in the spectrum which might
alleviate this situation. This goes under the notion  of assisted inflation
discussed in Refs.~\cite{assist1,assist2}.  We show that about ten of such
tachyonic fields can be sufficient enough  to inflate the three spatial
dimensions. This allows to generate adequate density perturbations.  We can
have more than one tachyon field  if we begin with a set of parallel and 
non-coincident $\tilde{D4}$ branes.  As explained earlier, this might occur if
there were many  pairs of $D5$ and anti $D5$ branes which yield many unstable 
non-BPS $D4$ branes.  Finally, these $\tilde{D4}$ branes  decay into
stable $D3$ branes. However, if we assume that these branes are non-coincident
and parallel then  the tachyon on a given brane  does not see the presence of
other tachyons. This is an essence of assisted tachyon condensation which we
shall stress upon.

We organize the paper by a brief discussion on effective action of BPS and
non-BPS branes.  Then we review the idea of assisted inflation and show that a
similar situation takes place in a system  where more than one tachyon fields are
present.  By requiring a consistent picture for inflation we obtain a bound on
the number of non-BPS branes, and show how  right amount of  density
perturbations is produced in such a context. Finally we comment on how
reheating takes place.


\section{Effective action of BPS and non-BPS brane}

In this section we briefly describe the world volume actions for BPS and
non-BPS branes which govern their dynamics. To keep the
description simple we ignore the fermions and concentrate only on
massless bosonic fields for the BPS branes, but include the tachyon field
for the non-BPS branes. The interested reader can consult Refs.~\cite{APS}, 
and~\cite{sennbd} for details. The world volume action for both BPS and
non-BPS branes is described by the sum of the Dirac-Born-Infeld (DBI) term
and the Wess-Zumino (WZ) term. The DBI action for a BPS $p$-brane is given by
\be
S_{DBI}^{(p)} = - T_p \int d^{p+1} \sigma e^{-\phi} 
\sqrt{- det (G_{\mu\nu} + F_{\mu\nu} )}\,,
\ee
where $T_p$ is the tension of the brane and 
$G_{\mu\nu} = G_{MN} \partial_\mu X^M \partial_\nu X^N$. Here $M,N$ are
the ten-dimensional indices taking values from $0,1,.....,9$; 
$\sigma^\mu(0\leq \mu \leq p)$ denotes the coordinates on the world volume of the
brane; while $G_{MN}$ is the ten dimensional background metric and 
$F_{\mu\nu}= \partial_\mu A_\nu - \partial_\nu A_\mu$ is
the world volume Born-Infeld field strength. We have ignored the
Kalb-Ramond 2-form field here for simplicity but the gauge-invariance of
the action demands its presence. The metric appearing in the above action is the
induced metric on the brane. The background metric $G_{MN}$ is not
arbitrary but restricted to satisfy the background field equations. The
transverse fluctuations of the D-brane is described by $9-p$ scalar
fields $X^i$ for $p+1 \leq i \leq 9$, and, the gauge field $A_\mu$ describes
the fluctuations along the longitudinal direction of the brane.

The WZ term for the BPS $p$-brane is a topological term and is given by
\be
S^{(p)}_{WZ} = \int d^{p+1} \sigma C \wedge e^{F}\,,
\ee
where the field $C$ contains Ramond-Ramond (R-R) fields and the leading term 
has a $p+1$ form. This acts as a source term for the brane and its presence is 
required for consistency of the theory like anomaly cancellation.

On contrary, the non-BPS $p$ brane has an extra tachyon field both in
the DBI and the WZ actions. The corresponding actions can be written down as
\begin{eqnarray}
S_{DBI}^{(p)} &=& - \int d^{p+1}\sigma~ e^{-\phi} 
\sqrt{- det (G_{\mu\nu} + F_{\mu\nu} )}\nonumber \\ & &  \qquad  
 \times F(T,\partial_\mu T, D_\mu\partial_\nu T)\,,
\, \nonumber \\
S^{(p)}_{WZ} &=& \int d^{p+1} \sigma C \wedge dT \wedge e^F\,,
\end{eqnarray}
where $T$ is the tachyon field and $F$ is some function of its arguments.
Its definition is such that for $T = 0, F = T_p$. The field $C$ in the WZ
action again contains the R-R fields but the leading term in $C$ is now a
$p$-form. Note, for a constant $T$, the WZ action simply vanishes and
for such a background the function $F(T,...) = V (T)$, where $V(T)$ is the
tachyon potential. On general grounds, it has been argued that at the
minimum $T_0$ of the potential $V(T)$ vanishes, i.e. $V (T_0) = 0 $. Thus at
$T = T_0$ the world volume action vanishes identically and in this case
the gauge field acts as a Lagrange multiplier field. This imposes a
constraint such that the gauge current also vanishes identically, i.e. forcing
all the states which are charged under this gauge field to disappear from the
spectrum.

However, the tachyon potential can be such that it admits a kink profile
for the tachyon field. This has been the most interesting issue recently.
A lot of attention has been drawn to the tachyon on an unstable non-BPS
$p$-brane, which condenses to form a kink and eventually forming a stable 
BPS $(p-1)$ brane. The kink solution for the tachyon is expected to give 
a $\delta$-function from $dT$-contribution, and, thus to 
reproduce the standard Wess-Zumino term for a resulting D$(p-1)$-brane. 
The kink solution effectively reduces the dimension of the world volume by one.

Although, finding out an explicit form for the tachyon potential is a difficult
proposition, however, string field theory predicts
an approximate form. This admits the kink formation besides preserving the
qualitative features. For example, it has been argued in Ref.~\cite{minahan} that 
the tachyon dynamics is described by the action given by
\be
 S = - \int dt~ d^p x \sqrt{-g} 
 \left({1\over 2} \partial_\mu T \partial^\mu T -  V(T)\right) ~,
 \label{action}
\ee
where the potential is written as
\be 
V(T) =  T^2 \ln T^2 + V_0 \,,
 \label{pot}
\ee
where we have defined the string scale to be unity $M_s=1$. The 
constant term; $V_0=e^{-1}$,  ensures that at the minimum located at 
$|T_0| = 1/\sqrt{e}$, one has $V(T_0) = 0$. See, Fig.~1.
We will assume this form of the potential in what follows.  The tachyon field
also couples to $U(1)$ gauge field living on the world-volume theory of the
non-BPS brane, see Ref.~\cite{minahan} for details. As we shall discuss below, 
for
our purpose we need more than one non-BPS brane. In such a case, if we
consider $n$ non-coincident but parallel non-BPS Dp branes, then the 
world-volume
theory will have $n$ tachyons which are neutral with respect to $(U(1))^n$ gauge
group. To be more precise, we are taking the separation between two
neighbouring branes is larger than the string length. In this picture, the
open string which begins from a given non-BPS D4 brane always ends on the
same brane and there are no open string stretched between two different
branes. Each of the open string which begins and ends on the same brane
has a tachyon in its spectrum.
Thus, the tachyonic action for such a system is a simple generalization of
what
we have written above for a single brane scenario, but where $T$ is to be
replaced by $T_i$ with a summation 
over $i$
running from $1$ to $n$. Condensation of these tachyons result a system of
$n$ non-coincident but parallel BPS D(p-1) branes.

\begin{figure}[t]
\centering
\leavevmode\epsfysize=4.4cm \epsfbox{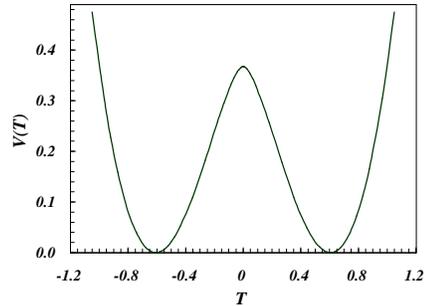}\\
\caption[fig1]{The shape of the symmetric potential in Eq.~(\ref{pot}). 
The height  is given by $V_0=e^{-1}$, while the minimum  occurs for 
$T_0=1/\sqrt{e}$. }
\end{figure}

\section{Tachyon dynamics}

Let us now address the problem of tachyon dynamics as governed by the action 
in Eq. (\ref{action}). Let us begin by remarking that the expression we are using to
model the profile of the potential has an unstable point. This is 
a false vacuum where the tachyon sits before condensation. 
Hereafter, we strictly assume that the tachyon is only the zero mode, it has 
no explicit dependence on any spatial dimensions.
When tachyon condensates it sits at the minimum of the potential.  In this
process, the tachyon rolls down the potential and gradually forms a
kink which becomes a stable $D3$ brane with a tension given by
\begin{equation}
\label{tension}
T_{3} = V_0+T_{4}\,,
\end{equation}
where $T_{3,4}$ are the brane tensions of $D3, \tilde{D4}$ branes. 
Notice, the tachyon being a homogeneous scalar field in our case 
can give rise a negative pressure. If the field is homogeneous within a 
patch of a Hubble size, then the patch may inflate the $3$ spatial dimensions. 
It turns out that a single tachyon system may not fulfill the required conditions, 
since it rolls down rather fast. 
However, if there is a set of unstable, parallel $\tilde{D4}$ branes separated 
from each other such that their tachyons do not interact among themselves,
then inflation may occur. In order to simplify our situation we make some
additional assumptions. One of the assumptions we have already mentioned and 
repeat here because of its importance. We treat the tachyon as a zero mode 
with a time dependence. This is related to the fact that presence of branes 
do not modify the underlying background space time structure.
This allows us to write down the background world volume metric as
\begin{equation}
ds_{p+1}^2=dt^2-a^2(t)(dx_1^2+dx_2^2+dx_3^2)- b^2 \sum_{i=4}^p dx_{i}^2\,,
\end{equation} 
where $a(t)$ is the scale factor along the three spatial dimensions. The
factor $b$ is  not time dependent and it defines the common size of the
extra spatial dimensions which we have assumed to be compactified on a torus. 
The actual value of $b$ is not crucial for our later discussion, although we  
take it to be of order of the fundamental string scale $(M_{\rm s})^{-1}$. 
Notice, that the metric is homogeneous and this apriori need not be the most
general  setup. However, we assume this for our purpose.  The above metric only
tells us that the branes are sitting on a space-time where there is no
backreaction on the metric due to their presence. These simple assumptions
allow us to understand the situation from $3+1$ dimensions.

Let us briefly comment on the mechanism which keeps the additional
dimensions stable. There are couple of points to be noticed, first 
of all the common size of the additional dimensions is small. Therefore, it is easy
to dynamically stabilize them. The fact that the radii are same leads to a
single {\em common radion mode} from an effective $3+1$ dimensional point of view.
The common mode has no dependence on the transverse spatial dimensions. 
The prescription of stabilizing the extra dimensions can be obtained 
dynamically following Ref.~\cite{abdel}.  Here we simply recall some of its 
features. From the point of view of $3+1$ dimensions it is possible to provide 
a time varying positive running mass to the common mode. If the three spatial
dimensions are expanding with a Hubble parameter $H$, then it has been observed 
in Ref.~\cite{abdel} that the  common mode actually gets a running 
mass of order $\sim {\cal O}(H)$. Once this happens,
the common mode decouples from the rest of the tachyon dynamics, and
settles down to the minimum of its potential. This stabilizes the 
extra compact dimensions during the very first few e-foldings of inflation,  
thus, here onwards we can safely  neglect their dynamics.

Let us now analyze in more details the dynamics of the tachyon as given by the
potential in Eq.~(\ref{pot}). This discussion is general and can be made in any
arbitrary dimensions. The only difference is a scaling factor that enters on
the potential, which shall
appear due to integrating out the additional spatial dimensions. 
Since, $M_{\rm s}$ is close to the Planck scale, this scaling
introduces only small factors which shall not  affect the actual
evolution of the tachyons. However, such a factor can play an  important role
while discussing density perturbations. For a moment we keep the discussion in
units where $M_{\rm s}=1$.

\begin{figure}[t]
\centering
\leavevmode\epsfysize=4.4cm \epsfbox{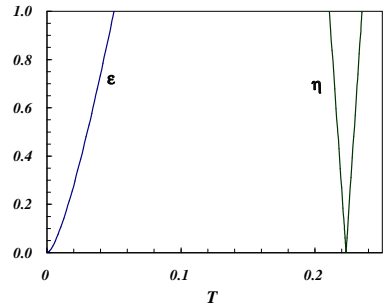}\\
\caption[fig2]{The two slow-roll parameters; $\epsilon$, and $\eta$, plotted 
with respect to the tachyon field $T$. Notice that 
the slow-roll conditions in Eq.~(\ref{slowroll1}) are not simultaneously  
satisfied  as the field rolls down 
from top of the potential to the global minimum.}
\end{figure}

As it has been observed in Ref.~\cite{burgess} the tachyonic  potential
is expected to have a profile similar to that of a Mexican hat potential;
$(T^2-T_0^2)^2$ which is indeed the case of the potential in Eq.~(\ref{pot}),
see Fig.~1. Whether inflation takes place or not can be verified very easily,
because most  of the inflationary models are based on slow-roll inflation,
which lays some simple conditions to be satisfied. In order to answer this
query we  study the inflationary conditions lead by the slow-roll parameters,
which  are given by
\begin{equation}
\label{slowroll1}
\epsilon \equiv  \frac{1}{2}\left(\frac{V^{\prime}}{V}\right)^2 \ll 1\,,
\quad \quad \quad \eta \equiv  \left|\frac{V^{\prime \prime}}{V}\right| \ll 1\,,
\end{equation}
where prime denotes derivative with respect to the field. As expected the above
conditions are not fulfilled at the same time during  roll over of the field
within a range $0 \leq T \leq T_0 \equiv 1/\sqrt{e}$. Indeed, one can easily
check that at the top of the potential, $\epsilon$  is close to zero, because
$V' \approx 0$.  However, $\eta$ is much larger than one.  Since $V''(T) = 4\ln
T + 6$, $\eta$ has a logarithmic divergence for $T\sim 0$, and remains large
until the effective mass term for the potential vanishes at $T=e^{-3/2}$.  On
the other hand, as $\epsilon$ increases gradually, both parameters  together
never become less than one simultaneously, and thus  the potential in
Eq.~(\ref{pot}) never provides a slow-roll inflation. 
This has been depicted in Fig.~2, where we have plotted $\epsilon$ and 
$\eta$. The field value where $V^{\prime \prime}$ vanishes is 
known as a spinodal  point because  at that point 
the mass squared term in the potential  flips its sign from 
negative to positive. The conclusion is that the tachyon rolls down very fast 
and this was precisely the observation made already in Ref.~\cite{burgess}.

For our potential inflation indeed takes place but for large values of 
$|T| >1$ ~\cite{barrow}, but this does not serve the present case. This rules out 
any possibility of having inflation with a single tachyon field.  Alternative
ideas have been discussed to cure this. Usually, it is believed  that inflation
might take place via altering the inter brane  separation~\cite{burgess,gia}.
In this paper we suggest a completely  new way to address this problem. We
suggest that inflation can be produced by a set of tachyonic fields,
all with the same potential as given by
Eq.~(\ref{pot}). Such a scenario can  appear from a system involving
many non-interacting, and non-coincident parallel $\tilde {D4}$ branes in type
IIB string theory. One of the key ingredients which we shall use here is
the non-interacting property of the tachyons. Eventhough, these tachyons are
not coupled to each other,  they are coupled dynamically via the expansion of
the Universe. This is an  essence of assisted inflation which was originally
discussed in  Ref.~\cite{assist1}. Thus, the question we are interested in is
asking whether we can have assisted inflation with a multi-tachyon
configuration. This is the issue we shall study in coming sections.

\section{Assisted inflation}

\subsection{Toy model}

In order to motivate the idea, we start by briefly recalling the main 
features of assisted inflation as originally presented in Ref.~\cite{assist1}.
Let us begin with a set of exponential potentials of the form 
\begin{equation}
V_i(\phi_i)  =   V_0 \exp \left( \alpha_i \, \phi_i \right) \,,
\end{equation}
where each scalar field has an exponential potential with a different slope
$\alpha_{i}$. 
Notice that the fields are not directly
coupled, albeit the combined role of the fields affect the expansion rate of the 
Universe:
\begin{eqnarray}
\label{Motion}
H^2 & = & \frac{8\pi}{3m_{{\rm p}}^2} \sum_{i=1}^n \left[ V_i(\phi_i) +
\frac{1}{2} \dot{\phi}_i^2 \right] \,, \\
\label{eqofmo}
\ddot{\phi}_i & = & - 3 H \dot{\phi}_i - \frac{dV_i(\phi_i)}{d\phi_i} \,,
\end{eqnarray}
where $H=\dot{a}/a$ is Hubble's constant, $m_{\rm p}$ is the Planck scale,
and $a$ is the scale factor of the flat FRW Universe in $3+1$ dimensions. 
Had there been a single field with 
an exponential potential, the solution would have been a power-law
solution 
\begin{equation}
\label{power}
a(t)  \propto  t^{p}\,, \quad \quad \quad p =\frac{16\pi}{m_{\rm p}^2\alpha^2} \,,
\end{equation}
where $\alpha$ is assumed to be an arbitrary slope. Notice, that the  solution
is inflationary only if $p>1$, i.e. for extremely shallow exponentials; $\alpha
\ll 1$. This suggests that if the potential is very steep, the field would roll
down fast and there would be no inflation at all. Now let us imagine that all
the fields  have such a steep slopes and  all of them are contributing to the
dynamics, then the modified $\tilde p$ is given by
\begin{equation}
\label{slope}
\tilde p = \frac{16 \pi}{m_{{\rm p}}^2} \sum_{i=1}^n \frac{1}{\alpha_i^2} \,.
\end{equation}
If we assume that all the slopes are the same then $\tilde p = np$. This
suggests  that potentials with $p<1$, which for a single field are unable to
support inflation, can do so as long as there are enough scalar fields to make
$np > 1$. This means that more the scalar fields, the quicker is the
expansion of the Universe. This is the reason the authors in
Ref.~\cite{assist1} have named such a cumulative phenomena as assisted
inflation. A generalization of this has been studied in a subsequent paper in
Refs.~\cite{assist2}, where various other potentials have been considered.
There is a clear message from both the
papers which we need to bear in mind that the dynamics of the assisted
inflation works well only when the fields do not have any explicit couplings
between themselves. Any kind of coupling tends to kill the assisted nature and
this is the sole criterion which we also have to fulfill.
There is another important reason behind choosing exponential potentials,
because there exists a late time attractor solution for all the participating fields, see
Refs.~\cite{assist1,karim}. This property is important while
calculating the density perturbations generated by the scalar fields exactly.
Lack of this behavior leaves the existing density perturbation calculations 
under some doubt.


\subsection{Our case}

Our case is not very different from the analysis of the above toy model,  
but for the nature of the potential. As we have already discussed in
section II if we consider a configuration of $n$ parallel non-BPS D4
branes which are separated from each other by a distance more than the
string length, then there is a tachyon in the world volume theory of each
brane and they donot interact with each other. The total potential from
the tachyons for this configuration is just the sum of the 
potential from each tachyon and is given by
\begin{equation}
\label{pot2}
V = \sum_{i=1}^n V_i = \sum_{i=1}^n\left(T_{i}^2\ln T_{i}^2 + V_0\right)\,,
\end{equation}
where $n$ is the total number of tachyonic fields, or, in other words the total
number of unstable non-BPS $D4$ branes, or, as a final configuration;
the number of stable $D3$ branes. Notice, that the individual tachyons are as 
such unaware of each others presence except that they all support the Hubble
expansion.  Thus they are all in a dynamical contact. The inflationary scale
has been implicitly assumed to be the string scale $\sim M_{\rm s}$. Due to our
choice in the compactification the string
scale can be related to the Planck scale by the compactification volume as
$M_{s}^{2+n}\cdot vol \approx m_{\rm p}^2$.

The foremost point to notice is the following; for the potential 
Eq.~(\ref{pot2}), the slow-roll conditions are now different. They are given by
\begin{equation}
\label{slowroll3}
\epsilon_i \equiv  \frac{1}{2}\left(\frac{1}{V} 
{\partial V\over \partial T_i}\right)^2  \ll 1\,,
\quad \quad \quad 
\eta_i \equiv  
\left|\frac{1}{V}{\partial^2 V\over \partial T_i^2}\right| \ll 1\,,
\end{equation}
As now the potential is a summation of all the tachyons
present in spectrum, this changes the behavior of the slow-roll parameters.
Our task is to ensure that they are both satisfied together while all $T_{i}$
roll down their respective potential. This can be seen very easily;
let us assume for the time being that 
$T_{1}(t)\sim T_{2}(t)\sim,...,\sim T_{i}(t)\sim T(t)$, this means that all the 
trajectories are moving on a single line on a phase space diagram. This we will
show  later on. If we assume so, then the total potential becomes $V=n V_1$, 
and the slow-roll conditions  become
\begin{equation}
\label{slowrn}
\epsilon_i \sim \left(\frac{1}{n^2}\right)\epsilon\,, 
\quad \quad \quad \eta_i \sim \left(\frac{1}{n}\right)\eta\,,
\end{equation}
where $\epsilon$ and $\eta$ are given as for a single field case,
\be
 \epsilon =  {1\over 2} 
 \left[{2 T\left(\ln T^2 + 1\right) \over T^2 \ln T^2 + V_0}\right]^2~;
 \qquad
 \eta = {|\ln T^4 + 6|\over T^2 \ln T^2 + V_0  }~.
 \ee
 
Next, our objective is to obtain the minimal number of branes required  to
obtain a region in the field space where $\eta_{i}$ and $\epsilon_i$ can be
made smaller than one in order to satisfy the slow-roll conditions. Notice,
that around $T=0$,  $\eta_i$ still has a logarithmic divergence.  Therefore, we
do not expect that the slow-roll conditions be satisfied right on top of
the tachyon potential.  However, as the fields roll down their respective
potential, there exists a  domain where $\epsilon_i$ can be made less than one
along with $\eta_{i}$. The valid region holds good until $T=e^{-3/2}$, the
point where $\eta$ goes to zero. Therefore, if we demand that
$\epsilon_i(e^{-3/2})\ll 1$, we can estimate the number of branes needed to
suppress $\epsilon$, and, $\eta$. This gives
 \be 
n\gg\sqrt{\epsilon(e^{-3/2})}= {\sqrt{8}\over \sqrt{e} \left( 1 - 3
 e^{-2}\right)}\approx 2.89~.
 \ee
For instance, if one requires $\epsilon_i<0.1$ around the spinodal point
$T=e^{-3/2}$, we will need a larger number of non-interacting $\tilde{D4}$ 
branes. In fact the number of branes turns out to be $n\geq 10$.

One can check the above statements by plotting  $\epsilon_{i}$, and $\eta_{i}$
with respect to  an arbitrary $T_{i}$, and looking  for the  region in field
space where both the slow-roll  parameters are less than one. In order to
illustrate this point we have  plotted in Fig.~3 the slow-roll parameters  as
given by Eq.~(\ref{slowrn}), for  $n=10$.  The large number  of fields
add up in the potential which ultimately affects the Hubble expansion
of the Universe. As a result the slow-roll conditions Eq.~(\ref{slowroll1}) are
satisfied simultaneously for  field values ranging from  $0.1 < T_{i} < 0.3$.
As we increase the number of fields, the overlapping region widens up. This is
the desired feature of assisted inflation which we wanted  in our case.
Assisted inflation by multi-tachyon really helps to inflate our  three spatial
dimensions, which are the longitudinal direction of all the $D3$ branes system.
This is a unique feature which shows an intrinsic characteristic of
all the branes. We do not require an uncertainty of moving the branes in a
higher dimensional space time. Even if the branes have a slow relative motion,
our scenario can be realizable. All it requires is that the tachyons developing
an individual kink does not communicate with others. We should mention that
last condition can be easily realized as long as  the inter brane separation is
larger than the string scale and the branes remain parallel to each other. 
In this paper we do not  intend to
find out an exact form of the scale factor, which we do not actually require
for the density perturbation calculation. Finally, let us note here the
versatile nature of the occurrence of the potential Eq.~(\ref{pot}). The
potential  can occur in one-loop supersymmetric corrections to the masses in a
bosonic sector, which gives a running mass to the field. This is a common
feature while studying  inflation in supersymmetric theories.

\begin{figure}[t]
\centering
\leavevmode\epsfysize=4.4cm \epsfbox{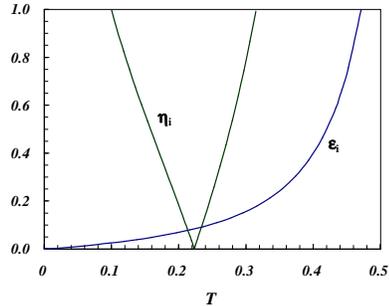}\\
\caption[fig3]{The two slow-roll parameters; $\epsilon_{i}$, and $\eta_{i}$ are plotted 
with respect to a single tachyon field $T_{i}$ when the total number of 
tachyons is taken to be $n=10$. Notice, now the slow-roll 
conditions can be satisfied concurrently in the region where
the tachyonic fields are rolling down their respective potentials.}
\end{figure}

Inflation apart from  making the Universe very flat also generates density
perturbation causally.  What actually matters is the last number of e-foldings
before the end  of inflation. This is because the observable Universe has to be
well within  the horizon at the time of density perturbations are being
produced. The number of e-foldings thus required is a model dependent issue,
which involves the uncertainty of two important scales; one at which 
inflation takes place, and the other, the temperature at which  the Universe becomes
radiation dominated. In our case we fix the inflationary scale to be the string
scale and in the next section we delve into the reheating temperature of the
Universe.

Before we move onto that we briefly mention the dynamics  of the
tachyon fields. Notice, that the equations of motion for the tachyons in $3+1$
dimension follow a simple relationship:
\begin{equation}
\frac{\dot T_{i}}{\dot T_1}=\frac{T_{i}}{T_1}\frac{1+2\ln T_{i}}{1+2\ln T_1}\,,
\quad \quad \quad {\rm for ~~~i \neq 1}\,,
\end{equation}
where dot denotes derivative with respect to time. While deriving the above
equation we have assumed the slow-roll conditions to be valid for all the
tachyons. We notice, that in the interesting regime  $|\ln T_{i}| \gg 1/2$, the
above expressions can be easily rewritten in a form
 \be 
 {d\over dt}( \ln\left(\ln T_i\right))= 
 {d\over dt}( \ln\left(\ln T_1\right))~.
 \ee
The above relationship is very important. It ensures that if the slow-roll
conditions are  satisfied, then there can be an unique solution, which tells us
that all  the tachyons follow a similar trajectory with an unique late time
attractor, such that 
\begin{equation}
\label{attract}
T_{1}(t) \sim T_{2}(t)\sim \,, ...\,, \sim T_{i}(t)\,.
\end{equation}   
This is an interesting  result, because other than the exponential potentials,
such a late time attractor behavior is very rare. In our case, it holds for a
wide region  of phase space, all that it requires is that during the slow-roll
inflation all the tachyons must follow $|\ln T_{i}| \gg 1/2$, which is very
genuine.  This allows us to perform the density perturbation calculation
without  any doubt.


\section{Density perturbations and reheating}

In this section we estimate the amplitude of the density perturbations
produced  during inflation, and, the reheating temperature of the Universe.
While generating the density perturbations the tachyon trajectories follow 
Eq.~(\ref{attract}). We pursue our calculation of density perturbation
following  Refs.~\cite{stewart,assist1}, which depends crucially on the late
time attractor behavior of the fields. The Hubble expansion in our case is a
cumulative effect of all the stable BPS $D3$ branes which can be expressed as
\begin{equation}
\label{hub}
H^2 \approx \frac{8\pi}{3m_{\rm p}^2}n{V_{eff}}~,
\end{equation}
where we considered the potential contribution alone by assuming the slow-roll
conditions Eq.~(\ref{slowroll1}) are met.  Here, we have also assumed there are
$n$ tachyons with the same effective $3+1$ dimensional potential $V_{eff}$,
which has to be calculated after integrating out the transverse dimensions in
Eq.~(\ref{action}). The effect is a scaling on the potential such that
$V_{eff}= vol\cdot V$. By assuming that the transverse directions
are compactified on a torii,  we can obtain the
effective potential in $3+1$ dimensions as
\be 
 V_{eff}(T) = M_s^2
 \left[T^2 \ln \left({T\over m_{\rm p}}\right)^2+ m_{\rm p}^2 V_0\right] \,.
 \label{veff}
\ee

The amplitude of the perturbations can be calculated before the end of
inflation  while the tachyons are rolling down towards their respective minima.
Since, tachyons follow the trajectory given by Eq.~(\ref{attract}), we  can
directly follow the  results obtained in Refs.~\cite{assist1,stewart}. We
obtain the spectrum of the density perturbations is given by
\begin{equation}
\label{pert}
\delta_{H}(k)\approx {1\over 5\pi} {H^2\over \sqrt{n}~ |\dot T|} = 
{3 n\over 5\pi}~ 
	{V_{eff}^{3/2}\over m_{\rm p} |V'_{eff}|} ~,
\end{equation}
where one has to evaluate the right hand side at the moment when the
perturbations are leaving the horizon, which one can roughly considered to be 
the critical point $T\sim e^{-3/2} m_{\rm p}$. This
assumption restricts the tachyon trajectories before the last $50-60$
e-foldings of inflation as we have mentioned above. This is what counts for 
the observable Universe. Our estimation then gives
\be
\delta_{H}(k)\approx 
 \left({3 n M_s\over 20\pi m_{\rm p}}\right)
  \left(1 - 3 e^{-2}\right)^{3/2} 
\sim 2\cdot 10^{-2}\left(n M_s\over m_{\rm p}\right).
\label{pert2}
\ee
The overall amplitude of  the density perturbations is constrained by COBE,
which predicts the right hand  side of Eq.~(\ref{pert2}) to be  $\sim 10^{-5}$.
Therefore, in order to be consistent with a COBE result, we need to constrain 
either the fundamental scale $M_{\rm s}$, or, the number of tachyons required.
The lower the ratio $M_{\rm s}/m_{\rm p}$ is, the 
higher  the required  number of non-coincident unstable branes.   
If we suppose  $n\geq 10$ as needed for inflation, 
then the string scale is constrained; we get
$M_{\rm s} \sim 10^{(-2,-3)} m_{\rm p}/n\leq 10^{(-3,-4)} m_{\rm p}$.  
Our naive estimation suggests that if the string scale is around  the 
Grand Unification scale, then the required number of unstable branes can 
be about ten.

All the tachyons while inflating the Universe reach a particular point on the
field space where the slow-roll conditions break down. This obviously depends  
on the number of tachyons we have  in the spectrum, see Eq.~(\ref{slowroll3}).
Once, inflation ends, the era of  entropy production prevails. This is the
first natural step among the  post-inflationary  epochs. The Universe is
required to be heated up to provide a thermal bath with a radiation dominated
Universe. In the conventional inflationary models it is believed that the
classical  energy stored in the inflaton potential is converted into the
kinetic energy  of the radiation bath through which one can easily estimate the
final  reheat temperature.  The situation in our case is similar to the
traditional one. The final scenario of reheating may go as follows. After the
tachyons have rolled down the potential, they begin oscillations around the
minimum. Notice, the following; the tachyon forms a kink in order to form a
stable $D3$ brane. After forming the kink the tachyon has to completely decay. 
Part of its  energy density in a region where the $D3$ brane has formed
goes into the brane tension.  However, rest of the vacuum energy, that
originally out of the kink has to be released in a way that produces a  thermal
bath in the bulk. This happens while tachyons oscillate around their minimum.
The oscillations lead to a time variation  in the width of the kink. This in
turn  leads to a vacuum  instability. This phenomena has a similar feature as
a  parametric reheating of the Universe, where the coherent oscillations  of
the classical vacuum produces quanta in a non-thermal way. In our case, the
massless zero mode gauge fields are excited, possibly  along with the massive
gauge fields  appearing due to compactification of the spatial  dimension.

The main conclusion is that the total energy stored in the 
region where the tachyon has gone to the minimum can be released into 
exciting the gauge fields, while the 
energy in the region where the kink has formed shall be
translated into the brane tension following Sen's conjecture. The final reheat
temperature can be  estimated if we assume that the tachyons only couple to the
gauge fields  with some gauge coupling $\sim \alpha_{s}$, same for all
tachyons, then an approximate reheat temperature can be estimated by
\cite{inf} 
\begin{equation}
T_{rh} \approx 0.1\sqrt{\alpha_{s}^2 M_{\rm s}m_{\rm p}}\,.
\end{equation}
While writing the above expression we have however assumed that the tachyon
mass is determined by its vacuum configuration which is given by the string
scale  $\sim M_{\rm s}$. This might leads to a very high reheat
temperature $T_{rh} \leq \alpha_{s}\cdot 10^{16}$ GeV. Such a high reheat
temperature might not be acceptable on many other cosmological grounds, 
which warrants another phase of late inflation in order to dilute any 
unwanted species. A late decay of those heavy 
modes produced by the tachyon may also help to modify the above relationship 
bringing $T_{rh}$ much smaller.

\section{Conclusion}

Let us summarize our results.  Starting from the observation  that a non-BPS
$D4$ brane can decay into a stable $D3$ brane via  tachyon condensation, we
imagine this process as a dynamical one which might give rise to some
interesting cosmological phenomena. Then we have studied a single tachyon case
to show that  inflation is not possible while the tachyon is rolling down the 
potential because the usual potential supported from the string field theory
comes out to be too steep. The two slow-roll conditions are not simultaneously
satisfied and the tachyon rolls rather fast. However, as we have argued, if
there is  a set of parallel and non-coincident unstable non-BPS branes, then
tachyon  condensation  may lead to a successful inflation where both the
slow-roll conditions can be met simultaneously. This is a virtue of a multi
field dynamics where the tachyons do not interact among themselves but tied up
with the evolution of the Universe. In this setup the Universe can be  imagined
to be a set of parallel $D3$ branes. For the specific tachyonic potential which
we have considered here, we noticed that one requires at least more than three
non-BPS $D4$ branes  to  fulfill all our criterion and drive inflation. A  more
conservative point of view requires at least ten  such branes.  We have noticed
that while the tachyons are rolling down the potential they can generate
density perturbations which can match the observed COBE normalization.  The
final fate of the tachyons in each and  every brane is to form a kink,
identified as the final BPS $D3$ brane whose tension is given by the tachyon
energy trapped with the kink.  The left  over vacuum where the kink does not
form releases  its energy while the tachyon is oscillating around the true
minimum, thus reheating the world. Our estimation towards the reheat
temperature is quite large.  This is because the gauge fields which are coupled
to the tachyons have a string coupling. The tachyon condensates at a value
close to the string scale giving rise to  massive gauge fields. The massive
gauge fields also decay and they are all responsible for generating further
entropy, that may help to reduce our current estimation of the reheating
temperature.

\acknowledgements
The authors are thankful to Bin Chen, Feng-Li Lin, and Partha Mukhopadhyay
for discussions. A. M. acknowledges the support of {\bf The Early Universe network} 
HPRN-CT-2000-00152.



\begin{references}

\bibitem{inf}
A. Guth, Phys. Rev. D {\bf 23}, 347 (1981); 
A. D. Linde, {\em Particle Physics and Inflationary Cosmology}, 
           Harwood Chur (1990); 
E. W. Kolb and M. S. Turner, {\em The Early Universe}, Addison--Wesley,
        Redwood City (1990); 
A. R. Liddle and D. H. Lyth, {\em Cosmological Inflation
  and Large-Scale Structure}, Cambridge University Press (2000).

\bibitem{bunn}
E. F. Bunn, D. Scott and M. White, Ap. J. {\bf 441}, L9 (1995);
E. F. Bunn and M. White, Ap. J {\bf 480}, 6 (1997).


\bibitem{binetruy} 
P. Binetruy and M. K. Gaillard, Phys. Rev. D {\bf 34}, 3069 (1986).

\bibitem{banks} 
T. Banks, hep-ph/9906126; hep-th/9911067

\bibitem{brustein}
R. Brustein and P.J. Steinhardt, Phys. Lett. B {\bf 302}, 196  (1993); 
T. Banks, D. Kaplan and A. Nelson, Phys. Rev. D {\bf 49}, 779 (1994);
B. de Carlos, J.A. Casas, F. Quevedo and E. Roulet, Phys. Lett. B 
{\bf 318}, 447 (1993).


\bibitem{polchi}
J. Polchinski, {\em String Theory}, Cambridge University Press (1998).

\bibitem{sen1}A. Sen, JHEP{\bf 9912}, 027 (1999).

\bibitem{sen2}
O. Bergman and M. R. Gaberdiel, Phys. Lett. B {\bf 441}, 133 (1998).

\bibitem{sen3}
A. Sen, hep-th/9904207.

\bibitem{extra}
I. Antoniadis, Phys. Lett. B{\bf 246}, 377 (1990);
I. Antoniadis, K. Benakli and M. Quir\'os,  Phys. Lett. B{\bf 331}, 313 (1994);
K. Benakli, Phys. Rev. D{\bf 60}, 104002 (1999); 
     Phys. Lett. B {\bf 447}, 51 (1999);
N.Arkani-Hamed, S. Dimopoulos, and G. Dvali, Phys. Lett B {\bf 429}, 263 (1998);
     Phys. Rev. D {\bf 59}, 086004 (1999);
I. Antoniadis, N. Arkani-Hamed, S. Dimopoulos, and G. Dvali, 
       Phys. Lett. B {\bf 436}, 257 (1998); 
L. Randall and R. Sundrum, Phys. Rev. Lett.  {\bf 83}, 4690 (1999); 
L. Randall and R. Sundrum, Phys. Rev. Lett. {\bf 83}, 3370 (1999).


\bibitem{lyth}D. H. Lyth, Phys. Lett. B {\bf 448}, 191 (1999);
G. Dvali and S. H. H. Tye, Phys. lett. B {\bf 450}, 72 (1999); 
N. Kaloper and A. Linde, Phys. Rev. D {\bf 59}, 10130 (1999); 
A. Mazumdar,  Phys. Lett. B {\bf 469}, 55 (1999); 
N. Arkani-Hamed, S. Dimopoulos, N. kaloper and J. March-Russell, 
         Nucl. Phys. B {\bf 567}, 189 (2000);
R. N. Mohapatra, A. P\'erez-Lorenzana and C. A. de S. Pires,
      Phys. Rev. D {\bf 62}, 105030 (1999).

\bibitem{lukas}
A. Lukas, B. A. Ovrut and D. Waldram, Phys. Rev. D {\bf 61}, 023506 (2000); 
P. Binetruy, C. Deffayet and D. Langlois,  Nucl. Phys. B {\bf 565}, 269 (2000);
P. Binetruy, C. Deffayet, U. Ellwanger and D. Langlois,  
         Phys. Lett. B {\bf 477}, 285 (2000); 
R.N. Mohapatra, A. P\'erez-Lorenzana and C.A. de S. Pires, 
            Int. J. Mod. Phys. A {\bf 16}, 1431 (2001). 
A. Lukas and D. Skinner, hep-th/0106190. 

\bibitem{abdel} 
A. Mazumdar and A. P\'erez-Lorenzana, Phys. Lett. B {\bf 508}, 340 (2001).


\bibitem{alexander}
S. Alexander, hep-th/010503.

\bibitem{burgess} 
C. P. Burgess, M. Majumdar, D. Nolte, F. Quevedo, G. Rajesh and 
R.-J. Zhang, hep-th/0105204.

\bibitem{gia}
G. Dvali, Q. Shafi and S. Solganik, hep-th/0105203; 
G. Shiu and S.-H. Tye, hep-th/0106274.

\bibitem{turok}
J. Khoury, B. A. Ovrut, P. J. Steinhardt and N. Turok, hep-th/0103229; 
     hep-th/0105199; hep-th/0105212; 
K. Enqvist, E. Keski-Vakkuri and S. Rasanen, hep-th/0106282.

\bibitem{kallosh} 
R. Kallosh, L. Kofman and A. Linde, hep-th/0104073; hep-th/0106241. 



\bibitem{assist1}
A. R. Liddle, A. Mazumdar and F. E. Schunck, Phys. Rev. D {\bf58},
061301 (1998).  

\bibitem{assist2}
E.J. Copeland, A. Mazumdar and N.J. Nunes, Phys. Rev.D {\bf 60}, 083506 (1999);
A. M. Green and J. E. Lidsey, Phys. Rev. D {\bf 61}, 067301 (2000).


\bibitem{APS}
A. Sen, JHEP {\bf 9910}, 008 (1999).


\bibitem{sennbd}
E.A. Bergshoeff, M. de Roo, T.C. de Wit, E. Eyras and S. Panda,
JHEP {\bf 0005}, 009 (2000).


\bibitem{minahan}
J. A. Minhan and B. Zwiebach, JHEP {\bf 0009}, 029 (2000);
 JHEP {\bf 0010}, 045 (2000).


\bibitem{barrow}
J. D. Barrow and P. Parsons, Phys. Rev. D {\bf 52}, 5576 (1995).
 


\bibitem{karim}
K. A. Malik and D. Wands, Phys. Rev. D {\bf 59}, 123501 (1999).

\bibitem{stewart}
M. Sasaki and E. D. Stewart, Prog. Theor. Phys. {\bf 95}, 71 (1996).
 


\end{references}
\end{document}